\documentclass[a4paper]{aa}
\usepackage{graphicx}
\usepackage{txfonts}

\usepackage{url}
\usepackage{graphicx}
\usepackage[hyphenbreaks]{breakurl}
\usepackage{float}
\usepackage{subfig}
\usepackage{multirow}
\usepackage{amsmath}
\usepackage{hyperref}

\outer\def\gtae {$\buildrel {\lower3pt\hbox{$>$}} \over 
{\lower2pt\hbox{$\sim$}} $}
\outer\def\ltae {$\buildrel {\lower3pt\hbox{$<$}} \over 
{\lower2pt\hbox{$\sim$}} $}

\newcommand{\tess}{\sl TESS}

\begin{document}

\title{HiPERCAM and {\tess} observations of the rapidly rotating M7V star LP 89--187}
\author{Gavin Ramsay\inst{1}, J. Gerry Doyle\inst{1}, Stuart Littlefair\inst{2}, V. S. Dhillon\inst{2,3}, David Garcia Alvarez\inst{3,4}}

\authorrunning{Ramsay et al.}
\titlerunning{HiPERCAM observations of LP 89--187}
\institute{Armagh Observatory and Planetarium, College Hill, Armagh, BT61 9DG, N. Ireland, UK\label{inst1}\and
Astrophysics Research Cluster, School of Mathematical and Physical Sciences, University of Sheffield, Sheffield S3 7RH, UK\label{inst2}\and
Instituto de Astrofísica de Canarias, E-38205 La Laguna, Tenerife, Spain\label{inst3}\and
GRANTECAN, Cuesta de San José s/n, E-38712, Breña Baja, La Palma, Spain\label{inst4}\\
\email{gavin.ramsay@armagh.ac.uk}}

\date{}

\abstract {The discovery of a significant number of rapidly rotating
  low mass stars showing no or few flares in {\tess} observations was
  a surprise as rapid rotation has previously been taken as implying
  high stellar activity. Here we present {\tess} and HiPERCAM
  $u_{s}g_{s}r_{s}i_{s}z_{s}$ observations of one of these stars LP
  89--187 which has a rotation period of 0.117 d. {\tess} data
  covering three sectors (64.6 d) only show three flares which have
  energies a few $\times10^{33}$ erg, whilst HiPERCAM observations,
  which cover 0.78 of the rotation period, show no evidence for flares
  more energetic than $\sim10^{31}$ erg. Intriguingly, other surveys
  show LP 89--187 has shown weak H$\alpha$ in emission. We compare the
  flare energy distribution of LP 89--187 with low mass stars in the
  $\beta$ Pic moving group, which have an age of $\sim$24 Myr. We find
  LP 89--187 has a lower flare rate than the $\beta$ Pic
  stars. In addition, we find that TRAPPIST-1 analogue stars,
  which are likely significantly older than the $\beta$ Pic stars,
  show fewer flares with energies $>10^{33}$ erg in {\tess} data.  We
  examine the relationship between amplitude and period for a sample
  of low mass stars and find that more rapid rotators have a higher
  amplitude.}


\maketitle

\section{Introduction}

It is well known that as stars age, their rotation period declines
\citep{Skumanich1972}. Over the last few decades, we also know that as
stars age they also become less `active' (see
\citet{Davenport2019}). Stellar activity can manifest itself in
different ways; for example as starspots, narrow optical line
emission, X-ray emission and flare activity.  Although flares have
been seen on stars with early spectral-types, they appear more common
on low mass dwarf stars and fully convective
stars (later than M3/4V) in particular \citep{Pettersen1989}.

The launch of {\tess} in April 2018 allowed nearly month-long
photometric observations with 2-min cadence for tens of thousands of
stars \citep{Ricker2015}. Data from the first three months of the
mission revealed a sample of low mass stars (M0-M6V) which showed a
modulation on a period $<$1 d \citep{Doyle2019}. These Ultra Fast
Rotators (UFRs) are amongst the very fastest rotating main sequence
stars. Surprisingly, those UFRs with periods $<$0.3 d showed no or few
flares in their light curves. \citet{Gunther2020}, who used data taken
from the first two months of the mission, also found there was a
`tentative' decrease in the flare rate for stars with P$<$0.3 d.
Given that fast rotation is generally taken to imply a strong magnetic
field through the dynamo effect (e.g. \citet{HartmannNoyes1987}),
these findings were unexpected.\\

A systematic search for UFRs using all the southern ecliptic 2 min
cadence {\tess} data was reported by \citet{Ramsay2020}. Out of 9887
stars with T$<$14 mag (the {\tess} passband, $\sim$600-1000 nm) and
close to the main sequence, 609 were found to be low mass stars with a
period $<$1 day. Out of these, only 288 showed at least one flare. For
stars with periods $>$0.4 d, 51\% of stars are flare active, whilst
for stars with periods $<$0.2 d the fraction is 11\%. This strengthens
the findings of \citet{Doyle2019} and \citet{Gunther2020}.

Several studies have attempted to address why some UFRs show very low
flare activity.  \citet{Doyle2022} used the VLT to make
spectropolarimetric observations of ten UFRs and found that five had a
magnetic field around $\sim$1--2 kG (this is typical of low mass stars
\citep{Reiners2022}). It is therefore unlikely that the star's low
flare activity is caused by a low magnetic field strength. In another
study, \citet{Ramsay2022} used spectroscopic observations from the
Nordic Optical Telescope (NOT) to determine if UFRs showing no flare
activity were in binary systems.  They found that out of 29 targets
whose spectra were obtained over 3 nights, only one showed evidence
for clear radial-velocity variations indicating most showed no
evidence for being in binary systems.

Another possibility, investigated by \citet{Doyle2022} and
\citet{Ramsay2022} is related to an effect called super-saturation,
(e.g. \citet{Jeffries2011}) where magnetic field lines are not able to
reconnect since they extend beyond the stellar corona due to
centrifugal stripping as the Keplerian corotation radius moves inside
the X-ray emitting coronal volume. An alternative solution could be:
(i) flares on these stars are numerous but have a low energy and
therefore cannot be observed in the {\tess} pass band, or (ii) it
could be explained via the radiative transfer work by
\citet{Houdebine1996}.  The latter showed that at very high transition
region pressures, corresponding to active dMe atmospheres, the
chromosphere becomes a very efficient radiator at continuum
wavelengths, particularly in the UV/blue region. This is similar to
what we see in the blue continuum of intense solar flares, except that
in super-saturated M dwarfs, a large fraction of the stellar
chromosphere experiences intense heating deep in the atmosphere due to
the high transition region pressure. In contrast, in the solar case
only the flare kernels (i.e. the location in the chromosphere where we
see flare emission) are enhanced. Based on older 1D models, the flare
kernel was assumed to be due to electron beams, while newer 3D models
\citep{Druett2024}, do not require beams, either electrons or
ions. Both of these possibilities can be investigated by looking for
flux variability.

\begin{figure*}
    \centering
    \includegraphics[width = 0.9\textwidth]{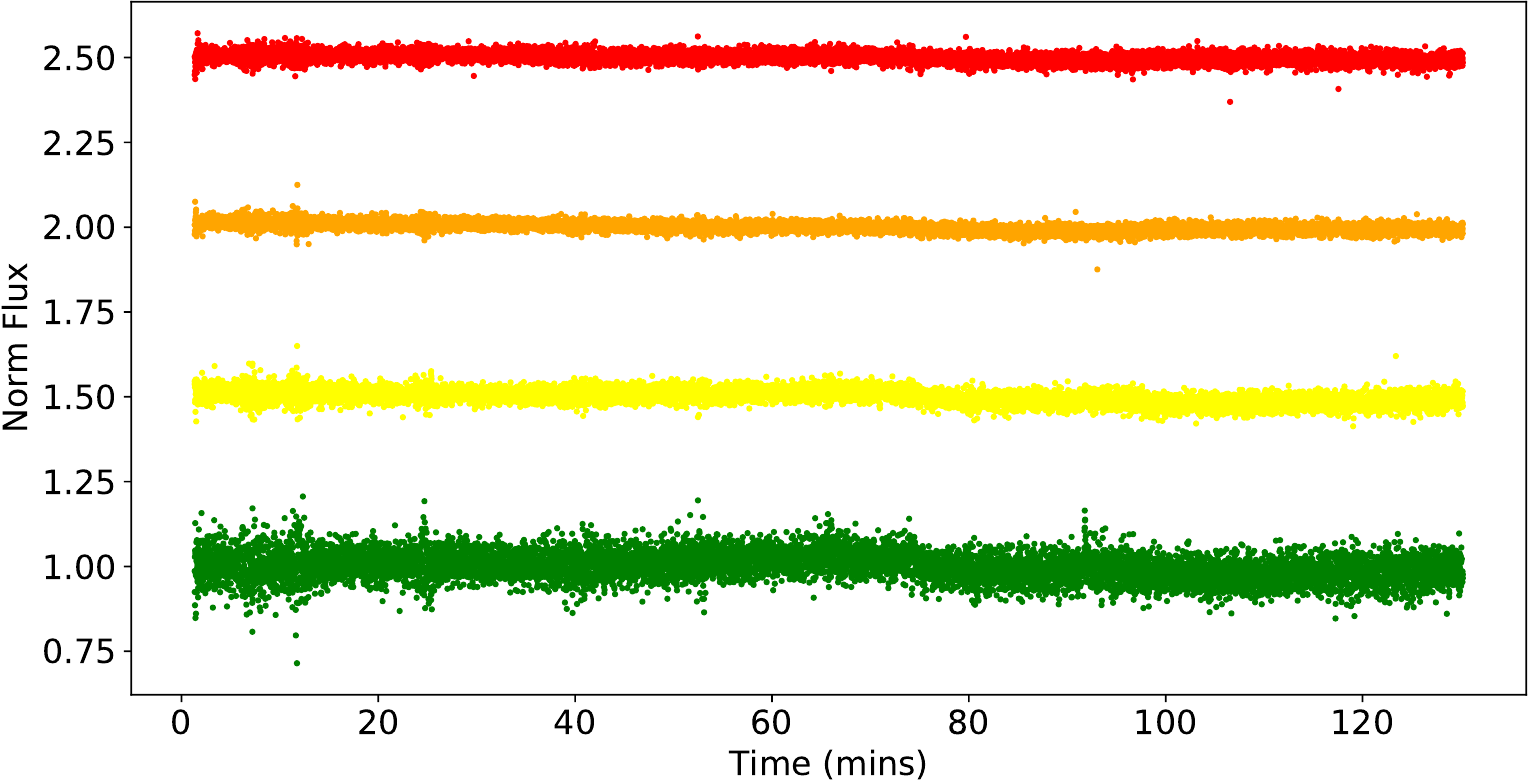}
    \caption{Differential $g_{s}$ (colour coded green), $r_{s}$
      (yellow), $i_{s}$ (orange) and $z_{s}$ (red) photometry of LP
      89--187 obtained using HiPERCAM. The differential flux for each
      0.74 s exposure has been normalised by dividing by the mean in
      each band and each subsequent filter has been shifted vertically
      by 0.5. Time zero corresponds to MJD=60350.902. The residual
      features, including the flare-like features in the $g_{s}$ band,
      are due to poorer conditions. The signature of the stars
        rotation has largely been removed due to the large colour
        difference between the target and the two comparison stars.}
    \label{hipercam-light}
\end{figure*}

In this paper we report on a search for short duration or low energy
flares using HiPERCAM from one rapidly rotating star, LP 89--187,
which showed no flares in the first sector of {\tess} data. We
  also constrain the rate of higher energy flares using an additional
  two sectors of {\tess} data.

\section{LP 89--187}

LP 89--187 was included in the \citet{Luyten1979} catalogue of high
proper motion stars. Subsequently Gaia data \citep{Gaia2021} indicated
it has a distance of 22.9 pc (see Table \ref{gaiaparameters} for the
key Gaia parameters). \citet{Newton2016} using MEarth photometric
observations found a short period modulation (0.117 d). LP 89--187
(TIC 80859893) was first observed in {\tess} Sector 20 in 2 min
cadence mode. These observations confirmed the short period modulation
(0.1165 d). In addition, this sector of data showed no evidence for
optical flares.  However, \citet{GarciaSoto2023} obtained 3 spectra of
LP 89--187 during a survey of low mass stars and determined an
H$\alpha$ EW of -5.25 \AA, indicating at least some evidence of
activity. Using LAMOST spectra, \citet{Wang2022}
determined a spectral type of M7 implying LP 89--187 is fully
convective (no evidence for Lithium was found in the spectrum).

We conclude that the photometric modulation is the signature of the
rotational period of the star for the following reasons. With a Gaia
{\sl RUWE} value of 0.958 (Table \ref{gaiaparameters}), it is not
likely to be in a binary system with a much more massive component
\citep{Castro-Ginard2024}. The $v sin i$ value of 66.0$\pm0.5$ km
s$^{-1}$ \citep{GarciaSoto2023} is entirely consistent with a rapidly
rotating star and the phase folded {\tess} light curves (Sect. 5) are
consistent with a star showing spots but not with binary systems which
show ellipsoidal modulation. Although we can expect some variation in
a stars photometric period over time due to spot migration, the spread
in the period of LP 89--187 derived from the three {\tess} sectors is
very low (see \citet{Ramsay2024} who show that low mass and solar type
stars can show a very large spread in the period derived from many
{\tess} sectors which is likely due to the detected period not being
the signature of the rotation period).  In the unlikely event that the
period of LP 89--187 was not the signature of the rotation period, but
rather a binary orbital period, then we would still expect the
component stars in binaries with periods shorter than $\sim$4 d to be
synchronised with the orbital period \citep{Lurie2017,Fleming2019}.
Based on these characteristics, we decided to observe LP 89--187 using
the high speed imager HiPERCAM.

\begin{table}
  \caption{Key parameters of LP 89--187 as determined by Gaia
    \citep{Gaia2021}.}
  \begin{center}
  \begin{tabular}{lr}
    \hline
    RA (2016.0) & 120.3378864 \\
    DEC (2016.0) & +56.4011159 \\
    Parallax (mas) & 43.64 \\
    $\mu_{RA}$ (mas/yr) & -20.00 \\
    $\mu_{DEC}$ (mas/yr) & -291.94 \\
    $G$ & 15.43 \\
    $BP-RP$ & 3.92 \\
    $RUWE$ & 0.958 \\
    \hline
  \end{tabular}
  \end{center}
  \tablefoot{Renormalised Unit Weight Error ($RUWE$) is a
    measure of the stars astrometric solution and can be used to
      determine whether it is a binary system.}
  \label{gaiaparameters}
  \end{table}

\section{Observations}

HiPERCAM is a multi-band optical imager mounted on the 10.4~m Gran
Telescopic Canaries (GTC) on La Palma. It obtains simultaneous images
in five bands ($u_s g_s r_s i_s z_s$) and can be read out at rates
faster than 1 kHz (see \citet{Dhillon2021} for details of HiPERCAM).

Observations of LP 89--187 were made using the GTC and HiPERCAM on
2024 Feb 10. The conditions were generally clear (although there was
some thin cloud at times), the mean seeing was $\sim1.2^{''}$
(although occasionally it got significantly worse, rising to over
$4^{''}$) and lunar conditions were dark. Images were obtained every
0.74 s in all five filters, with only 8 msec deadtime between
exposures. Two comparison stars were visible in the field of view. The
duration of the observations covered 2.18 hr, or 0.78 of the rotation
period of LP 89--187.

The data were reduced using the HiPERCAM data reduction pipeline
\citep{Dhillon2021}. Prior to aperture photometry, images were
de-biased and flat-fielded, whilst $i_s$ and $z_s$ images had also
fringe corrections applied. Aperture photometry of LP~89--187 was
carried out with apertures that varied with the measured full-width at
half-maximum in each frame to minimise the effects of seeing
variations on the photometry. The measured counts of LP~89--187 were
divided by the mean counts of the two reference stars to remove any
transparency variations. There is a large difference in colour between
LP 89-187 and the comparison stars. Secondary extinction effects make
the long term trend in the lightcurve unreliable and prevent detection
of the rotation signature, but do not affect the photometry on short
timescales and do not hamper our ability to detect flares.

\begin{figure*}
    \centering
    \includegraphics[width = 0.9\textwidth]{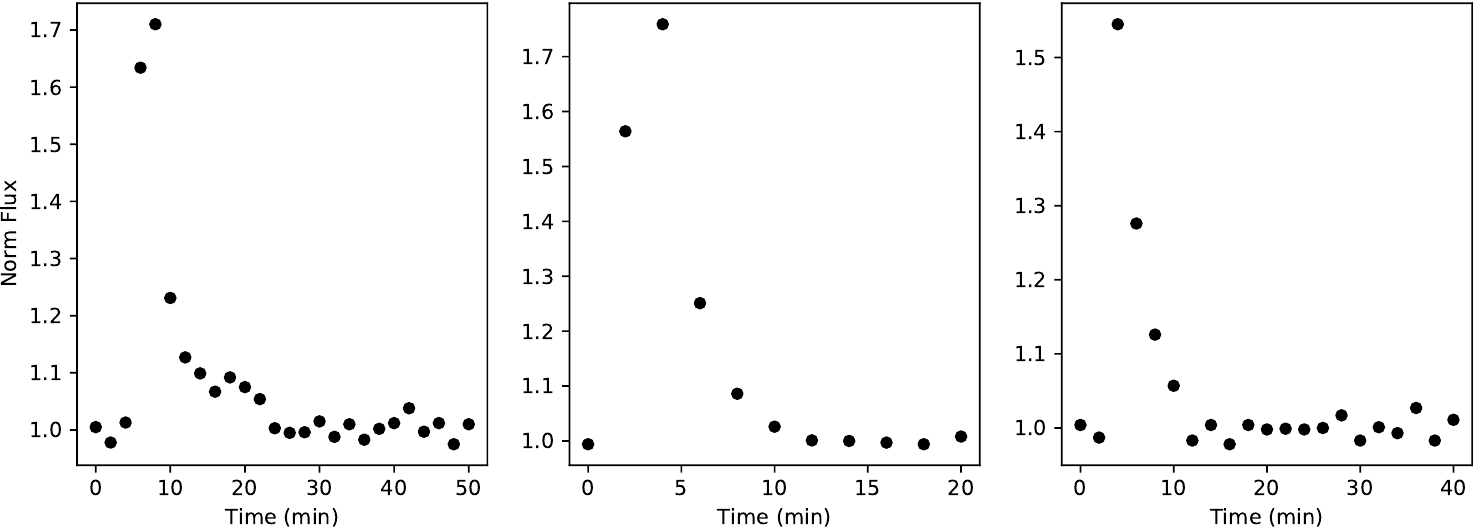}
    \caption{The three flares detected in {\tess} observations of LP
      89--187 using 2 min cadence data. The start time of each flare
      is, from left to right, MJD=59581.12, 59592.19 (both Sector 47)
      and 59942.01 (Sector 60).}
    \label{flare-light}
\end{figure*}

We show the light curve in the $g_s r_s i_s z_s$ bands in Figure
\ref{hipercam-light}: no stellar sources were detected in any of the
$u_s$ band images. The light curves show some residual features which
were caused by poorer conditions. There was no visual evidence for any
flares in any of the pass-bands. To make a more detailed search for
flares we used {\tt Altaipony} \citep{Davenport2016,Ilin2021}. For a
feature to be classed as a flare, three or more consecutive flagged
points were required. We did not find any flares in the $g_s r_s i_s
z_s$ images.

To estimate the luminosity of flares which we could have detected, we
again used {\tt Altaipony} to inject fake flares which had profiles
typical of stellar flares. For each $g_s r_s i_s z_s$ light curve we
created 500 fake light curves and injected flares with amplitudes in
the range 10$^{-8}$ -- 0.1 (i.e. a maximum of 10 percent above the
mean) and had durations in the range 0.7 - 30 min. We then searched
for flares in these fake light curves.

To determine the energy of these fake flares, we first obtained the
quiescent bolometric luminosity of LP 89--187 from the TIC V8.2
catalogue, $L$=0.00121 $L_{\odot}$ \citep{Paegert2021}.

We calculate the flare energy using:
\begin{equation}
\hspace{3cm}           E = L_{quiet} \times ED
\end{equation}
where $E$ is the flare energy, $L_{quiet}$ is the quiescent luminosity
of the star and $ED$ is the equivalent duration of the flare
\citep{Gershberg72} determined using {\tt Altaipony}.

However, given the temperature of flares are hotter than the
photosphere of a M7V star ($\sim$2700K), we have to take into account
that much of the flux will be outside the HiPERCAM
pass-bands. Inevitably there is uncertainty in the best choice of
temperature, with \citet{Howard2019} assuming 9000 K, but
\citet{Jackman2022} finding evidence for a temperature of $\sim$6000
K. We assume a flare temperature of 6000 K. (To convert the resulting
flare energies assuming a flare temperature of 9000K, we would
multiply these energies by a factor of 0.9, 1.3, 1.6 and 1.9 for $g_s
r_s i_s z_s$ respectively).  To determine a correction factor to
determine the bolometric flux for each passband, we convolved the
passband of the HiPERCAM filters with the blackbody (also an
approximation) of 6000 K to obtain correction factors: 15.4, 17.2,
24.6 and 49.7 for $g_s r_s i_s z_s$ respectively.  The minimum energy
for the fake flares in each passband were in the range
1.9$\times10^{31}$ to 5.2$\times10^{31}$ erg.

\section{Further {\tess} observations}

During our on-going analysis of low mass stars observed using {\tess},
we initially identified LP 89--187 as a rapidly rotating late type
star which showed no optical flares in 2 min cadence data obtained in
Sector 20. Since then, it has also been observed in Sectors 47 and 60,
again with 2 min cadence. We therefore searched the more recent data
for flares.

For each sector of data we first removed data points which did not
have {\tt QUALITY==0} flags, and then removed the signature of the
rotation period using {\tt kepflatten} which is part of the {\tt PyKE}
suite of tools\footnote{\url{https://pyke.keplerscience.org}}. To
search for and characterise flares we again used {\tt Altaipony}
\citep{Davenport2016,Ilin2021}, where we again required three or more
consecutive points to be flagged to be classed as a flare. This
inevitably implies that we cannot detect flares which have durations
shorter than a few mins. There were two consecutive events above the
local mean in sector 47 which could have been a short duration flare
although this could also have been due to noise. We note that from
{\tess} Cycle 5 there was the possibility of obtaining data with a 20
sec cadence.  We identified three flares and show their profiles in
Figure \ref{flare-light}. Each flare shows the typical stellar flare
profile: a rapid rise followed by a slower decline. The first flare
has a duration of $\sim$20 min with the other two a duration of
$\sim$10 min.

To determine the energies of the flares, we used the same procedure as
for the injected fake flares (see Sect. 3 and also
\citet{Ramsay2020,Ramsay2021}) using a correction factor of 3.1 for
the {\tess} filter where we assume a flare temperature of 6000 K. We
estimate the flares have an energy of 3.3, 2.7, 1.6$\times10^{33}$ erg
(left to right in Figure \ref{flare-light}). (In the unlikely
  event of LP 89--187 being a component of a binary system (see Sect. 2),
  then these energies would be reduced by a factor of two if the
  companion was, for instance, also a M7V star). The duration of the
observations is a combined total of 64.6 d. There is therefore a flare
with energy a few $\times10^{33}$ erg once every 21.5 d, or 1 per
0.0465 d$^{-1}$.

\section{Discussion}

We identified LP 89--187 as one of the low mass stars which rotates
very fast (0.117 d) but showed no evidence for optical flares during
its first sector of {\tess} observations. Given we usually expect such
rapid rotators to be active due to the dynamo effect, this is a
surprise. To investigate this further, we have examined additional
{\tess} data and obtained a short amount of high cadence photometry
using HiPERCAM mounted on the GTC.

\begin{figure}
    \centering
    \includegraphics[width = 0.45\textwidth]{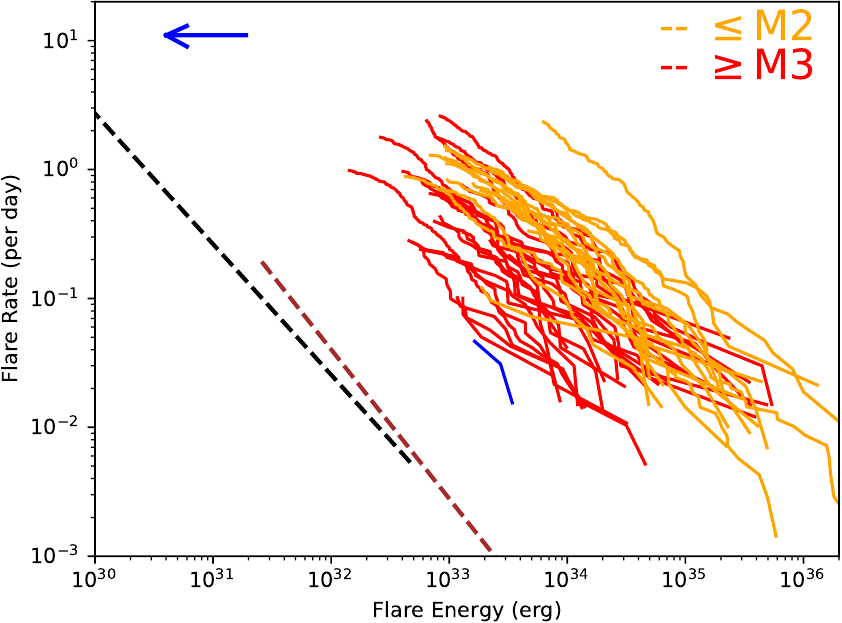}
    \caption{The bolometric flare frequency distibution of LP 89--187
      as derived from three sectors of {\tess} data shown in blue. The
      blue arrow indicates the upper limit derived from the HiPERCAM
      observations. We also show the bolometric flare frequency
      distribution of stars in the $\beta$ Pic moving group in orange
      (spectral type $\buildrel {\lower3pt\hbox{$<$}} \over
      {\lower2pt\hbox{$\sim$}} $ M2) and in red (spectral type
      $\buildrel {\lower3pt\hbox{$>$}} \over {\lower2pt\hbox{$\sim$}}
      $ M3). We also show flare frequency distribution of TRAPPIST-1
      analogue stars (black dashed line) and TRAPPIST-1 using {\sl K2}
      data (brown dashed line) taken from Figure 21 of
      \citet{Seli2021}.}
    \label{flare-rate}
\end{figure}

LP 89--187 was observed in three {\tess} sectors giving data covering
64.6 d. We detected three flares, all with energies $>10^{33}$
erg. This implies a flare once every 21.5 d or with a frequency of
0.0465 d$^{-1}$. We show the flare frequency distribution using the
{\tess} data and the upper limit derived from the HiPERCAM data in
Figure \ref{flare-rate}. To place these flare frequency distributions
in the wider context, we sought out a suitable sample of stars to
compare.

Determining robust ages for M dwarfs is difficult (see
\citet{Soderblom2010} for details).  However, stars in open clusters
or moving groups are particularly useful since the cluster age can be
determimed. One group of stars which has been studied by {\tess} is
those in the $\beta$ Pic moving group which has an age of 24$\pm$3 Myr
\citep{Ealy2024}. Although LP 89--187 has a proper motion which is not
consistent with that of this moving group, given its rapid rotation it
is likely to be a very young star. (\citet{Lu2024} determine a
gyrochronology relationship for partially and fully convective stars,
but this is not calibrated for stars younger than 0.67 Gyr).  However,
given the absence of Lithium in its spectrum \citep{Wang2022} it is
unlikely to be much younger than the $\beta$ Pic group (see
\citet{GutierrezAlbarran2024} for a recent study of the Lithium age
relation). There is some evidence that fully convective dwarfs
  can remain fast rotating for several Gyr, but slow down quickly
  after that \citep{Newton2016}. However, \citet{Pass2022} found a
  small number of low mass stars which appear to have spun down
  substantially by 600 My due to the nature of the stars magnetic
  field.

To ensure we compare like-with-like, we downloaded all of the
available {\tess} data of the 49 stars analysed by
\citet{Ealy2024}. Some of these stars have been observed in multiple
sectors, with the most recent being observed in Sector 71 (which ended
Nov 2023).  We again used {\tt Altaipony}
\citep{Davenport2016,Ilin2021} to identify flares in the {\tess} data,
where we removed the signature of rotation and ensured there were at
least three consecutive points in a flare. Similarly we assumed a
temperature of 6000 K and applied a correction to obtain the flare
bolometric luminosities. Of the 49 stars in the sample, seven did not
have a luminosity in the TIC 8.2 catalogue \citep{Paegert2021}, with
four of these being classed as giant stars: we did not determine the
energy of the flares for these stars.

We show the flare frequency distribution of LP 89--187 and the 42
stars in the sample of \citet{Ealy2024} in Figure
\ref{flare-rate}. The flare distribution of LP 89--187 indicates fewer
flares in the $\sim$\~10$^{33}$ erg range than the stars in the
$\beta$ Pic sample. The rotation period of these stars are in the
range 0.20 -- 8.74 d, implying that LP 89--187 is rotating more
rapidly than stars in the $\beta$ Pic sample. We note that there is
some evidence in Figure \ref{flare-rate} that stars with a spectral
type {\gtae} M3 show fewer flares than stars with spectral type
{\ltae} M2: this seems to be the reverse of the finding by
\citet{Ealy2024}. It is unclear why this is the case.

The HiPERCAM observations are of relatively short duration (2.18 hr)
and do not fully cover the rotation period. However, there was no
evidence for emission in the $u_s$ band in any of the individual
images. Further, there was no evidence of flares in the $g_s r_s i_s
z_s$ bands. By injecting fake flares to these light curves we are
confident that we would have detected flares with energies greater
than a few $10^{31}$ erg. If we extrapolate the flare frequency
distribution of the stars in the $\beta$ Pic sample to lower energy we
find they are likely to have higher flare rates at these less
energetic flares than inferred for LP 89--187. Extrapolating a
plausible flare frequency distribution of LP 89--187 to
$10^{31}$ erg, we estimate the HiPERCAM observations provide an upper
limit of 20 per day for flares with energies $>10^{31}$ erg, or
1.8 flares during the HiPERCAM observations.

We also compare these flare rates with stars which are close in the
colour-absolute magnitude diagram to TRAPPIST-1 which is an M8V star
hosting seven transiting exoplanets. \citet{Seli2021} analysed the
{\sl TESS} 30 min cadence data of 248 stars and identified 94
flares. The derived ages of the stars is very wide but have a peak in
their distribution at $\sim$2 Gyr (the 1$\sigma$ uncertainty is
typically 2--3 Gyr). In Figure \ref{flare-rate} we show the flare
frequency distribution of TRAPPIST-1 and its `analogue' stars obtained
using {\sl K2} data (Figure 21 in \citet{Seli2021}). The stars in the
young $\beta$ Pic moving group and LP 89--187 all show flares with
energies \gtae10$^{33}$ erg. (We reduced energy of the flares in
\citet{Seli2021} by 1.6 since they assumed a flare temperature of
9000K rather than the 6000K we have assumed). In contrast, the
TRAPPIST-1 analogue stars show flares with energies \gtae10$^{33}$ erg
at a rate which is $\sim$2 orders of magnitude less than the $\beta$
Pic stars -- this is expected because of their difference in
  age.

\begin{figure}
    \centering
    \includegraphics[width = 0.5\textwidth]{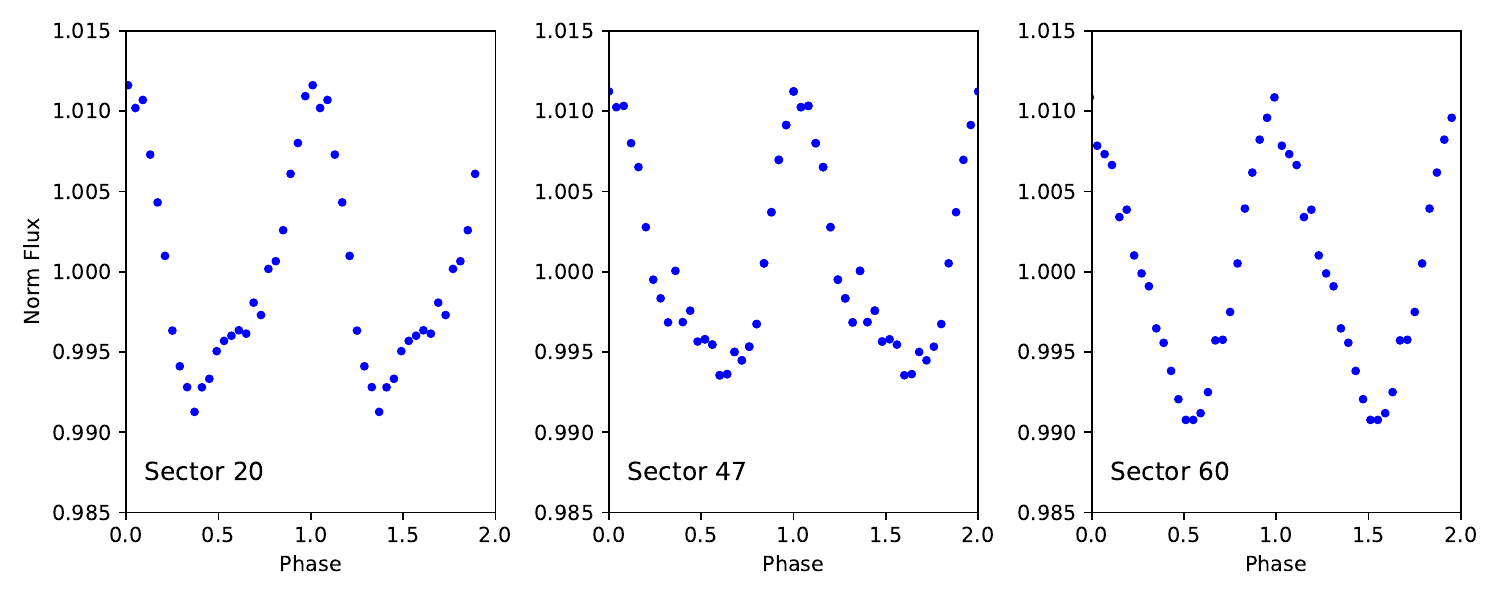}
    \caption{The folded light curves of LP 89--187 in the three
        {\tess} sectors. The phase has been shifted so that flux
        maximum is at $\phi$=0.0.}
    \label{rot-amp}
\end{figure}

We now compare the amplitude of the rotational modulation of LP
89--187 with other low mass stars, where we determine the peak-to-peak
amplitude of a Fourier series fit to the light curve at the dominant
period (we use Lomb Scargle periodogram as implemented in the {\tt
  VARTOOLS} suite of software \citep{Hartman2008}). The average
amplitude from 3 sectors is 1.655$\pm$0.102 \% and we show the
phase folded and binned light curves for the different sectors in
Figure \ref{rot-amp}.
  
\begin{figure}
    \includegraphics[width = 0.45\textwidth]{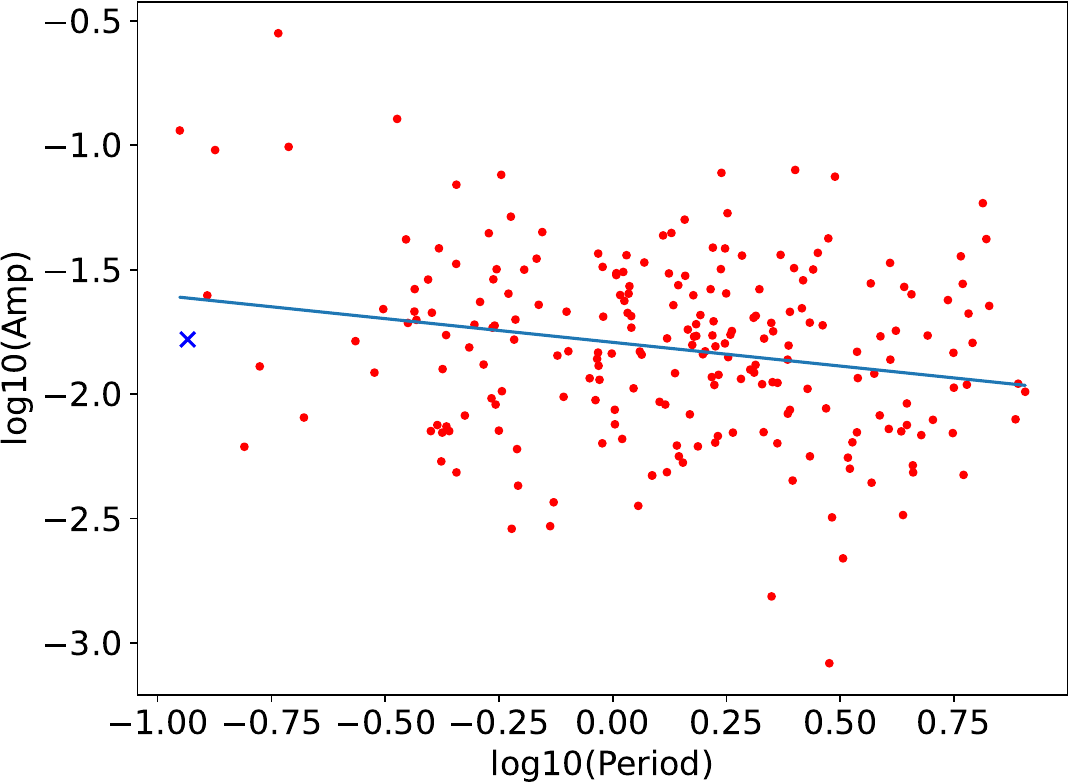}
        \caption{The amplitude of the modulation versus period of stars
          taken from \citet{Ramsay2024} which have a low spread in their
          period between {\tess} sectors ($\delta_{LS}$ < 0.05).  LP 89--187 is shown as a
          blue cross and the slope has an index of -0.19 and
          significance of $p$=0.0016.}
    \label{period-amp}
\end{figure}

How does this compare with other low mass stars? In \citet{Ramsay2024}
we examined {\tess} data of low mass stars in both the northern and
southern continuous viewing zones, implying they were observed in
multiple sectors. We were able to identify a sample of stars which had
a very low spread in their derived period between
sectors. Stars which had a higher spread in their period indicated that
the observed period was {\sl not} the signature
of the rotation period. For those stars with a low spread in their
period (where LP 89--187 lies), we show the amplitude of their
rotation as a function of period in Figure \ref{period-amp}. There is
a significant correlation ($p$=0.0016) between period and amplitude
with a slope of --0.19$\pm$0.06. This suggests that the inhomogeneity
of darker regions (magnetic activity) is greater for more rapidly
rotating stars: this would again be at odds with the relatively
low number of flares seen in LP 89--187.

We now briefly discuss one possible explanation for rapidly rotating
low mass stars showing relatively low rates of optical flares. In the
1990’s, the existence of polar spots was hotly debated,
e.g. \citet{Schussler1992} and references therein. Further work by
various authors (e.g. \citet{Holzwarth2007}, \citet{Isik2024})
suggested that high latitude spots on fast rotating M dwarfs is due to
the dominance of the Coriolos force. Spots at high latitude can
interact with open field structures leading to increased stellar wind
and reduced flare production. It is possible that these rapidly
rotating young stars have a dominant spot location at high latitudes
which leads to reduced flare activity. This is consistent with work
done by \citet{Hallinan2007,Hallinan2008} who reported intense radio
emission from the polar regions of a sub-set of rapidly rotating (a
few hrs) ultra-cool M dwarfs. This region is only visible on objects
whose axis is inclined towards Earth. The presence of such a region on
ultra-fast rotators can lower the amount of free energy available for
flares, hence a lower rate of flare activity.

\section{Conclusion}

Observations with Kepler and {\tess} have revealed a subset of UFRs
($P<$0.3 d) M dwarfs that have reduced flare activity. One of these
stars was LP 89--187 which has a 0.117 d rotation period. Observations
by \citet{GarciaSoto2023} showed the LP 89--187 has H$\alpha$ emission
at several epochs.  We have analysed three {\tess} sectors of data of
LP 89--187, detecting three flares with energies of $\sim10^{33}$
erg. We also used HiPERCAM to monitor for optical flares at lower
energy, failing to detect any flares over the course of the
  relatively short observation span. We also compared the flare rate
of TRAPPIST-1 analogue stars, which have a similar spectral type to LP
89--187, but older than the $\beta$ Pic group stars. We find that the
young $\beta$ Pic stars show flares with energies $>10^{33}$ erg at a
rate several orders of magnitude more frequent than the older
TRAPPIST-1 analogues. In contrast, LP 89--187, whose age is unclear,
but likely not much younger than the $\beta$ Pic stars, shows
$10^{33}$ erg flares but at a lower rate than the $\beta$ Pic
stars. Determining the age of LP 89--187 is of great interest:
  this maybe possible if we could identify it as belonging to a moving
  group of stars with known age.

\begin{acknowledgements}

VSD and HiPERCAM operations are funded by the Science and Technology
Facilities Council (grant ST/Z000033/1).  Based on observations made
with the Gran Telescopio Canarias (GTC), installed at the Spanish
Observatorio del Roque de los Muchachos of the Instituto de
Astrofísica de Canarias, on the island of La Palma, under program ID
GTC120-23B. The design and construction of HiPERCAM was funded by the
European Research Council under the European Union’s Seventh Framework
Programme (FP/2007-2013) under ERC-2013-ADG Grant Agreement no. 340040
(HiPERCAM). This paper also includes data collected by the {\tess}
mission whose funding is provided by the NASA's Science Mission
Directorate. JGD would like to thank the Leverhulme Trust for a
Emeritus Fellowship. Armagh Observatory \& Planetarium is core funded
by the N. Ireland Executive through the Dept. for Communities.

\end{acknowledgements}

\end{document}